# Magnetron Sputtered Non-Toxic and Precious Element-Free Ti-Zr-Ge Metallic Glass Nanofilms with Enhanced Biocorrosion Resistance

*Baran Sarac,\* Matej Micusik, Barbara Putz, Stefan Wurster,*

*Elham Sharifikolouei, Lixia Xi, Maria Omastova, Florian Spieckermann,*

*Christian Mitterer, Jürgen Eckert*

**Baran Sarac** – *Erich Schmid Institute of Materials Science, Austrian Academy of Sciences, 8700 Leoben, Austria; orcid.org/0000-0002-0130-3914; Email:* baran.sarac@oeaw.ac.at

**Matej Micusik** – *Polymer Institute, Slovak Academy of Sciences, Dubravsa cesta 9 Bratislava 84541, Slovakia*

**Barbara Putz** – *Erich Schmid Institute of Materials Science, Austrian Academy of Sciences, 8700 Leoben, Austria; Department of Materials Science, Montanuniversität Leoben, 8700 Leoben, Austria*

**Stefan Wurster** – *Erich Schmid Institute of Materials Science, Austrian Academy of Sciences, 8700 Leoben, Austria*

**Elham Sharifikolouei** – *Politecnico di Torino (POLITO), Department of Applied Science and Technology, Corso duca Degli Abruzzi 24, 10129, Turin, Italy*

**Lixia Xi** – *College of Material Science and Technology, Nanjing University of Aeronautics and Astronautics, Yudao Street 29, 210016 Nanjing, China*

**Maria Omastova** – *Polymer Institute, Slovak Academy of Sciences, Dubravsa cesta 9 Bratislava 84541, Slovakia*

**Florian Spieckermann** – *Department of Materials Science, Montanuniversität Leoben, 8700 Leoben, Austria*

**Christian Mitterer** – *Department of Materials Science, Montanuniversität Leoben, 8700 Leoben, Austria*

**Jürgen Eckert** – *Erich Schmid Institute of Materials Science, Austrian Academy of Sciences, 8700 Leoben, Austria; Department of Materials Science, Montanuniversität Leoben, 8700 Leoben, Austria*

The chemical composition and structural state of advanced alloys are the decisive factors in optimum biomedical performance. This contribution presents unique Ti-Zr-Ge metallic glass thin-film compositions fabricated by magnetron sputter deposition targeted for nanocoatings for biofouling prevention. The amorphous nanofilms with nanoscale roughness exhibit a large relaxation and supercooled liquid regions as revealed by flash differential scanning calorimetry. $Ti_{68}Zr_8Ge_{24}$ shows the lowest corrosion (0.17 µA $cm^{-2}$) and passivation (1.22 µA $cm^{-2}$) current densities, with the lowest corrosion potential of –0.648 V and long-range stability against pitting, corroborating its excellent performance in phosphate buffer solution at 37 °C. The oxide layer is comprised of $TiO_2$, $TiO_x$ and $ZrO_x$, as determined using X-ray photoelectron spectroscopy by short-term ion-etching of the surface layer. The two orders of magnitude increase in the oxide and interface resistance (from 14 to 1257 $\Omega$ $cm^2$) along with an order of magnitude decrease in the capacitance parameter of the oxide interface (from $1.402 \times 10^{-5}$ to $1.677 \times 10^{-6}$ S $s^n$ $cm^{-2}$) of the same composition is linked to the formation of carbonyl groups and reduction of the native oxide layer during linear sweep voltammetry.

## 1. Introduction

Due to their defect and grain-free amorphous structure and flexible chemistry, metallic glasses are nowadays regarded as potential candidates for various applications. Their exclusive combination of the high elasticity, mechanical strength, corrosion stability, and good resistance against hydrogen embrittlement renders metallic glasses for use as alternative materials for hydrogen storage, electrocatalytic converters, and lithium-ion rechargeable batteries.[1-7] Besides, due to their minimal stress shielding and anti-biocorrosion, biocompatibility with the promotion of cell adhesion, and in some cases, biodegradability, these advanced amorphous alloys have started to be used in daily healthcare applications such as orthopedics and dental implants, screws for bone fracture fixation, and stents for cardiovascular repair.[8-15]

Promising antimicrobial properties of thin film metallic glass (TFMG) coatings have been shown in the literature.[16-22] Biocompatible flexible metallic films were suggested to be used in smart windows, microfluidic channels, and antifouling coatings by adjusting the pattern of the wrinkled structure to achieve a dynamically tunable transmittance and wetting behavior [23]. Also, β-type Ti35Nb2Ta3Zr alloys increased the proliferation, alkaline phosphatase activity, calcium deposition and mRNA expression of MG63 osteoblast cells as compared to Ti6Al4V [24]. Another study has shown promising interfacial biocompatibility and osseointegration via anodic oxidation of Ti-24Nb-4Zr-7.9Sn (Ti 2448) alloy by analyzing the behavior of bone marrow stromal cells (BMSCs) cultured on the surfaces of Ti 2448 in vitro and performing post-histological analysis after in vivo implantation of the modified surfaces [25]. Furthermore, studies on metallic glass coatings reported low cytotoxicity upon exposure to C2C12 myoblasts,[26] L929 fibroblast cells,[27-29] MC3T3-E1 cells,[30] human mesenchymal stem cells (hMSC),[31] and murine (MC3T3-E1) and human osteoblast-like cells (SaOS-2),[32] as well as hydroxyapatite formation[33] and ensuring the minimal adhesion of cancer cells.[34] However, in all these studies, the Cu, Ni or Al used to enhance the glass-forming ability have the risk of being released as ions or metal from metal implant particles into the body tissue,[35-37] which could even kill sensitive cells in other parts of the body. Furthermore, studies have shown that a large fraction of Cu in the alloy (≥30 at.%) can initiate pronounced pitting events and hence must be avoided.[38, 39]

In order to enhance the glass-forming ability, the Ti-based metallic glasses discovered so far are composed of three or more elements.[40] A typical Ti-based glass former can be thus defined as $Ti_xM_yN_{100-(x+y)}$. Here, $M_y$ that can be a metal, metalloid or non-metal, typically having a large negative heat of mixing or a remarkable size difference with Ti. Despite the zero negative heat of mixing with Ti, Zr is an ideal candidate for this second element since the atomic size difference is 147 pm for Zr vs. 160 pm for Ti,[41] and it is non-toxic and abundant. From the elements with lower atomic numbers, Be has proven a great success in casting significantly large diameters of Ti-based MGs (above 50 mm) with its atomic radius 22.8% smaller than Ti.[42] As the third element, the highly toxic element Be can be replaced by B, Si, P or S, which have zero or negative heat of mixing with Ti. Among the metalloids, Ge can be a good candidate since $\Delta H_{mix\_Ti-Ge}$ = –51.5 kJ/mol and $\Delta H_{mix\_Zr-Ge}$ = –72.5 kJ/mol with a covalent radius of 121 pm.[41, 43] Furthermore, compared to Si, Ge has a significantly lower Young's modulus ($E_{Si}$ = 165 GPa, $E_{Ge}$ = 135 GPa),[44, 45] which is beneficial to eventually match with that of human cortical bone (~20 GPa).[46]

As a consequence of an unhealthy lifestyle and ignorance of oral hygiene, the number of cavities and teeth gum-related disorders, and in consequence, the demand for dental surgeries increases. Nevertheless, most dental implant materials currently used do not suffice the optimal longevity and durability either due to tissue-/osseointegration or biomechanical response. This contribution focuses on developing non-toxic and precious element-free Ti-based metallic glass nanofilm coatings targeted for dental and other biomedical applications. Three different compositions with a variation of the Ti to Zr content in the presence of Ge (21-25 at.%) were produced using magnetron sputter deposition and examined thoroughly using structural, thermal, morphological, compositional and electrochemical methods.

## 2. Results and Discussion

### 2.1. Structural, Thermal, and Optical Properties

The amorphous nature of the as-sputtered samples was checked using grazing incidence X-ray diffraction (**Figure 1a**). The broad amorphous diffraction maximum shifts from ~37.8° to ~38.6° as the amount of Ti in the TiNFs increases. Although the second broad hump is barely distinguishable for all the samples (between 65° and 85°), the determination of the peak position is not possible due to high scatter. The flash differential scanning calorimeter traces of the fabricated compositions display remarkable differences (**Figure 1b**). A small piece scraped from the thin film was inserted into the center of the high-sensitivity chip for the FDSC measurements (**Figure 1c**). Composition (3) – lowest Zr content - shows a clear shift towards higher temperatures. The relaxation temperature, $T_r$, is 706 ± 1 °C, the glass transition temperature, $T_g$, is 902 ± 1 °C, and the crystallization temperature is above the device limits ($T_x > 980$ °C). The relaxation ($T_g - T_r = 268 ± 2$ °C) and supercooled liquid (SCLR, $T_x - T_g = 116 ± 2$ °C) regions are largest for composition (2) – intermediate Zr content. Interestingly the SCLR of composition (2) is only 65 ± 2 °C. The results reveal that the decrease in the Zr content to 8 at.% leads to a pronounced increase in the thermal stability and probably the glass-forming ability. Table 1 summarizes the thermophysical properties of the developed TiNFs. The TiNFs sputtered on Si/SiO$_2$ show high reflectivity and a mirror-like surface (**Figure 1d**).

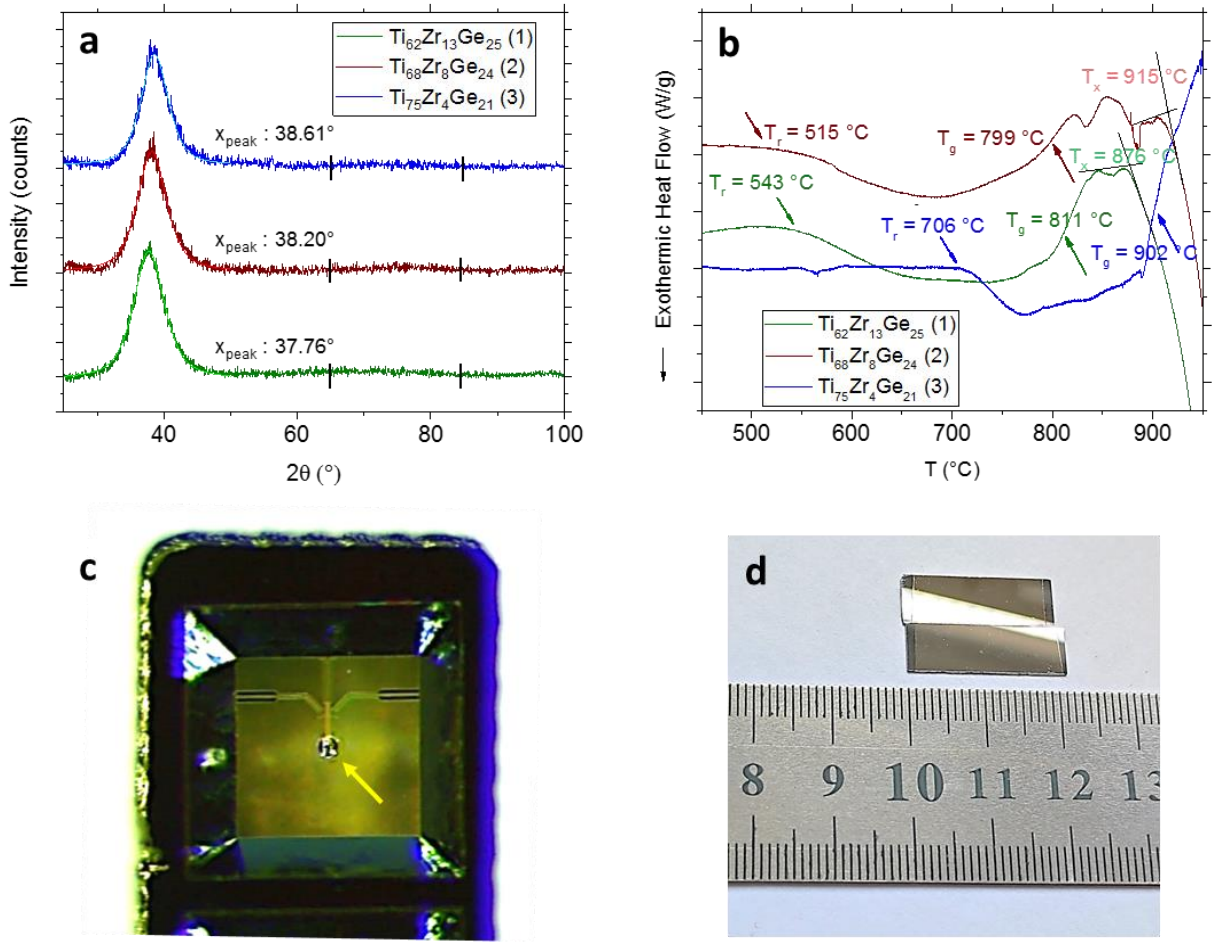

**Figure 1**. (a) Fully amorphous XRD patterns of the samples; the peak positions of the first broad maximum included were determined by a Pseudo-Voigt function. The regions for the second broad maximum were indicated by the black stripes. (b) FDSC traces (heating rate 250 °C min$^{-1}$) of the TiNF compositions. $T_r$: relaxation temperature, $T_g$: glass transition temperature, $T_x$: crystallization temperature. (c) A scraped thin-film piece (indicated by a yellow arrow) situated on the sensor part of the FDSC chip. (d) Magnetron sputtered TiNFs with glossy surfaces.

**Table 1**. FDSC data of the TiNF compositions. The errors of $T_r$, $T_g$, and $T_x$ are within 1 °C, whereas $T_g - T_r$, and $T_x - T_g$ have errors within 2 °C.

| Composition | $T_r$ (°C) | $T_g$ (°C) | $T_x$ (°C) | $T_g - T_r$ (°C) | $T_x - T_g$ (°C) |
|---|---|---|---|---|---|
| Ti$_{62}$Zr$_{13}$Ge$_{25}$ (1) | 543 | 811 | 876 | 268 | 65 |
| Ti$_{68}$Zr$_{8}$Ge$_{24}$ (2) | 515 | 799 | 915 | 284 | 116 |
| Ti$_{75}$Zr$_{4}$Ge$_{21}$ (3) | 706 | 902 | >980 | 196 | >78 |

## 2.2. Surface Tribology and Morphology

The results of topological investigations of the Ti-based TFMGs from the 3D AFM profiles obtained from the representative 5 × 5 μm² sections together with SEM surface images of similar areas are depicted in **Figure 2**. The largest root-mean-square roughness, $S_q$, along with the largest surface area, $A_{surf}$, is attained for composition (1) (**Figure 2a**), which could be due to the small features uniformly distributed throughout the surface (**Figure 2d**). On the other hand, the average height, $h_{avg}$, is the largest for composition (2) (**Figure 2b**), which means that the larger-scale surface irregularities are spread homogeneously throughout the sample, as corroborated by SEM imaging (**Figure 2e**). Composition (3) has the lowest values for these three parameters (**Figure 2c**), possibly due to inhomogeneously dispersed and fewer large surface asperities (**Figure 2f**).

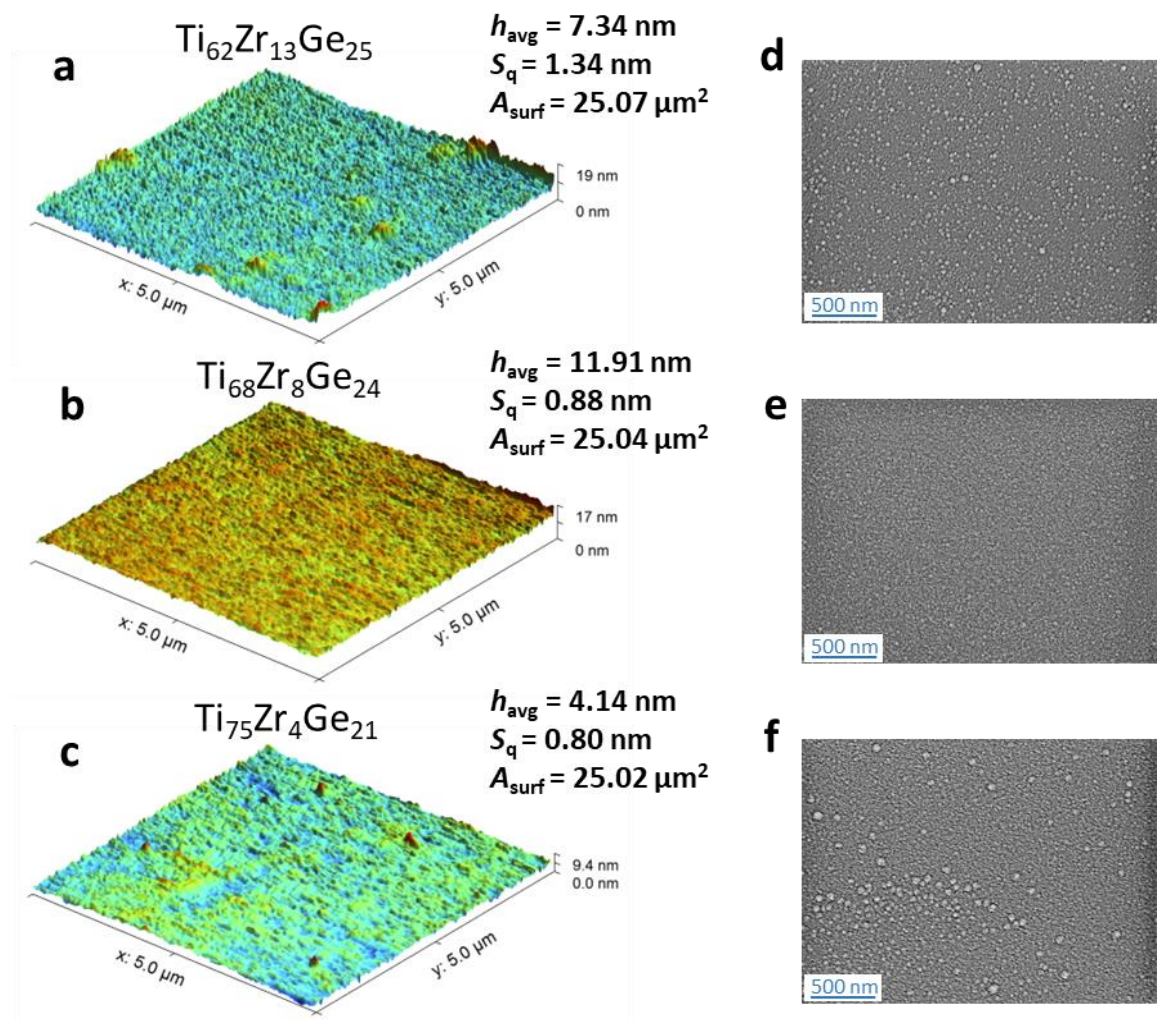

**Figure 2**. 3D AFM profiles of (a) $Ti_{62}Zr_{13}Ge_{25}$, (b) $Ti_{68}Zr_{8}Ge_{24}$, (c) $Ti_{75}Zr_{4}Ge_{21}$, and (d-f) their corresponding surfaces recorded by FESEM. $h_{avg}$: average height of surface asperities, $S_q$: root-mean-square roughness, $A_{surf}$: actual surface area.

## 2.3. Electrochemical Analyses - Experimental

The polarization behavior of the developed Ti-based compositions is investigated in **Figure 3a**. A decrease in the corrosion current density $j_{corr}$ is determined as the Ti content slightly increases at the expense of Zr (c.f. $j_{corr}$ = 225 nA cm$^{-2}$ for composition (1) and $j_{corr}$ = 166 nA cm$^{-2}$ for composition (2). However, further reduction of the Zr content significantly increases $j_{corr}$ to ~60 µA cm$^{-2}$. Hence, minor compositional adjustments have a dramatic impact on corrosion resistance. This phenomenon can be correlated to the glass-forming ability (GFA) which is in turn linked to the heat of mixing of components and atomic size. The predicted heat of mixings are 0 kJ mol$^{-1}$ for Ti–Zr, –51.5 kJ mol$^{-1}$ for Ti–Ge, and –72.5 kJ mol$^{-1}$ for Zr–Ge [43, 47]. Furthermore, when the atomic radii of these elements are compared, Zr (0.162 nm) is larger than Ti (0.147 nm), leading to a bigger atomic size difference with Ge (radius of 0123 nm) [41, 43]. Hence, the atomic bonding of Zr is stronger than with Ti, and the reduction of Zr in this system decreases the GFA [48]. This decrease probably also leads to the deterioration of the corrosion properties. The electrochemical findings are ~~also~~ compared with Oak et al.'s study performed under the same conditions (standard PBS solution at 310 K).[49] The typical passivation current density $j_{pass}$ of pure commercial Ti and Ti-6Al-4V alloys in PBS solution at 37 °C are on the order of 10$^{-5}$ A cm$^{-2}$. The developed compositions (1) and (2) have an order of magnitude smaller $j_{pass}$ values of 1.2–1.3 × 10$^{-6}$ A cm$^{-2}$, which are comparable to most of the developed Ti-based bulk compositions.[49] The most significant advantage of our thin films compared to the indicated Ti-based glasses is that they do not contain any toxic or noble-element (i.e., Cu and Pd). On the other hand, the composition (3) shows a significantly larger $j_{pass}$ indicating that further replacement of Zr with Ge leads to a remarkable difference in passive layer formation.

Another significant improvement is in pitting corrosion. The maximum pitting corrosion potential $E_{pit}$ observed for the Ti-Zr-Pd-Cu-Sn-(Ta,Nb) bulk rods is between 0.3 V - 0.6 V; furthermore, their cathodic corrosion potential $E_{corr}$ is around –0.1 V – 0.1 V.[49] Hence, a significant increase in $\eta_{pit}$, the difference between the pitting and cathodic corrosion potential,[50] can be attained by choosing adequate compositions within the Ti-Zr-Ge system. **Table 2** tabulates the findings from the potentiodynamic polarization curves. The repassivation potential $E_{rp}$ is obtained by reversing the potential from positive to negative.

We do not observe repassivation behavior for composition (3), which could be because the polarization experiment was reversed close to the onset of $E_{pit}$.

**Table 2**. Comparison of different Ti-Zr-Ge-based TiNF compositions. $β_c$: Cathodic beta, $β_a$: Anodic beta, $j_{corr}$: Cathodic corrosion density, $E_{corr}$: cathodic corrosion potential, $E_{pass}$: passivation potential, $j_{pass}$: passivation current density, $E_{pit}$: pitting corrosion potential, $E_{rp}$: repassivation potential, $η_{pit}$ (V) = $E_{pit} - E_{corr}$, $ΔE_{rp} = E_{rp} - E_{corr}$. Error: $E_{corr}$ ±0.005 V, $E_{pass}$ ±0.010 V, $j_{corr}$ ±0.02 μA cm$^{-2}$, $j_{pass}$ ±0.03 μA cm$^{-2}$, $E_{pit}$ ±0.007 V, $E_{pit}$ ±0.002 V.

| Compos. | $β_c$ (mV dec$^{-1}$) | $β_a$ (mV dec$^{-1}$) | $j_{corr}$ (nA cm$^{-2}$) | $E_{corr}$ (V) | $E_{pass}$ (V) | $j_{pass}$ (μA cm$^{-2}$) | $E_{pit}$ (V) | $E_{rp}$ (V) | $η_{pit}$ (V) | $ΔE_{rp}$ (V) |
|---|---|---|---|---|---|---|---|---|---|---|
| Ti$_{62}$Zr$_{13}$Ge$_{25}$ (1) | 63 | 62 | 225 | –0.522 | –0.350 | 1.28 | 0.790 | 0.153 | 1.312 | 0.503 |
| Ti$_{68}$Zr$_8$Ge$_{24}$ (2) | 113 | 146 | 166 | –0.648 | –0.390 | 1.22 | 0.565 | 0.075 | 1.213 | 0.465 |
| Ti$_{75}$Zr$_4$Ge$_{21}$ (3) | 537 | 102 | 62×10$^3$ | –0.176 | 0.245 | 8.64×10$^3$ | >0.80 | – | >0.976 | – |

The frequency-dependent behavior of the newly developed TiNFs is depicted in **Figure 3b-d**. The most considerable difference before and after LSV is observed for composition (3); the arc in the Nyquist plot becomes smaller after LSV. The impedance values, particularly $Z_{im}$, become larger for (1) and (2) after LSV, indicating higher stability on the surface layer. The maximum Bode angle determined are 86.2° and 85.9° for (2) and (1), respectively, at relatively low frequencies of ~1 Hz referring to a supercapacitive behavior in the PBS solution at 37 °C.[51] After polarization, composition (2) reaches the plateau at ~400 Hz at the highest |Z| of ~195 Ω cm$^{-2}$. For the as-sputtered sample, this value equals ~90 Ω cm$^{-2}$. The difference between the before and after polarization is not visible for composition (1), where the stabilization is reached at ~35 Ω cm$^{-2}$ and a frequency of ~1500Hz. |Z| is only ~5 Ω cm$^{-2}$ for composition (3) with a stabilization frequency of ~15000 Hz. For this reason, it can be deduced that the frequency-independent impedance can be reached much earlier for composition (2) because the ionic diffusion is relatively faster.[7, 52] This behavior reached at different frequencies was also shown for ZnCl$_2$ activated carbons carbonized at different temperatures [53]. The reason for the frequency-independent impedance is that at high frequencies, the Warburg impedance is quite small since the diffusing ions move only very little, where the resistance becomes independent of the applied frequency. On the other hand, at lower frequencies, the Warburg impedance increases since the ions have to diffuse farther [54]. Furthermore, the drop in the phase angle, from a high-capacitive behavior (~86.2°) to extremely low capacitance values is observed for composition (2) in a relatively smaller range of frequencies, which indicates the larger ionic resistance contributions that arise with pseudocapacitance as compared to double-layer capacitance [55, 56].

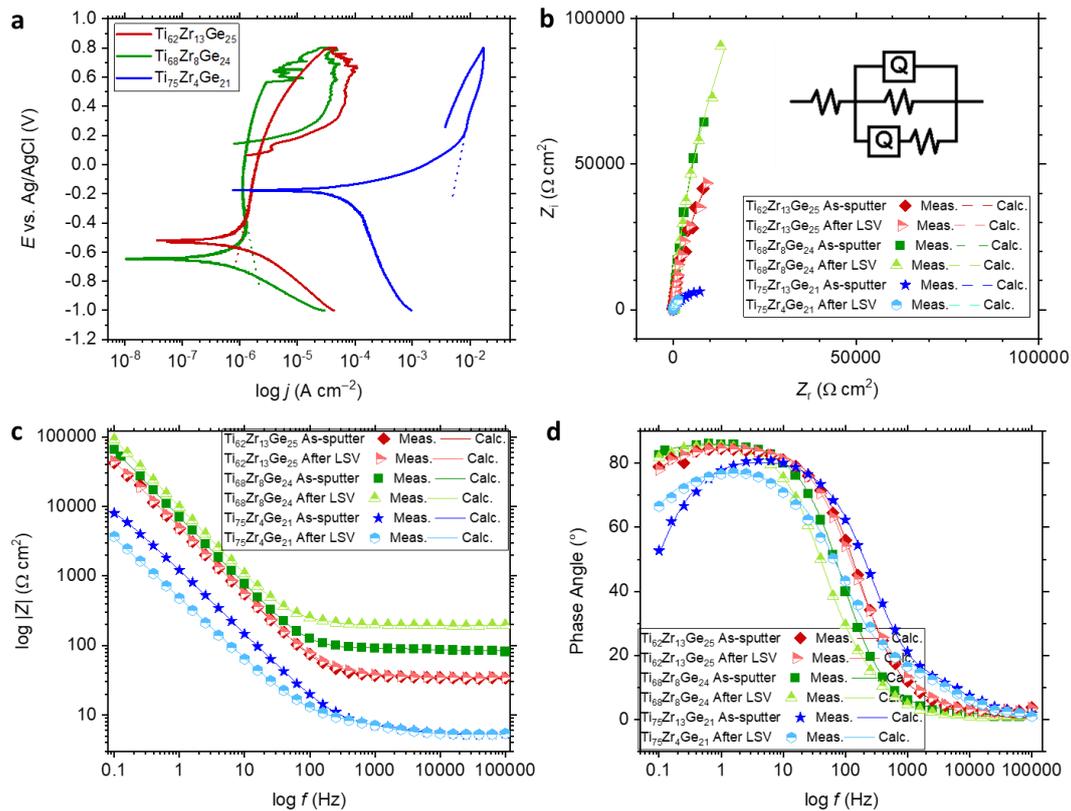

**Figure 3**. (a) Forward potentiodynamic polarization scans of the Ti-Zr-Ge compositions in PBS solution at 37 °C. The point where the dashed lines intersect with the curves indicates passivation onset. (b) Nyquist plots (Equivalent circuit model circuit shown in inset), (c) Bode phase and (d) Bode magnitude plots of the examined compositions at their open-circuit potentials in PBS solution at 37 °C.

## 2.4. Electrochemical Analyses – Modeling

A $R_s(Q_1R_1(Q_2R_2))$ equivalent circuit model (ECM) was proposed to simulate the EIS results. $R_s$ is the combination of contact resistance between the active electrode and its interface, electrolyte resistance, and internal resistance of active electrodes.[57] The parallel-connected $Q_1$ and $R_1$ indicate the constant phase element (CPE) for double-layer capacitance and charge-transfer resistance on the oxide surface, respectively. Here, $T$ and $n$ are the sub-components of $Q$ corresponding to the CPE parameter and exponent, respectively.[58, 59] In parallel to this connection, we have a series-connected $Q_2$ and $R_2$ related to the oxide layer and the interfacial interactions between oxide and metallic glass. $R_s$ of (1) and (3) remain almost constant after LSV. However, $R_s$ of (2) triples which is mainly due to the change in the contact resistance of the electrode upon fresh oxide formation. The $T_1$ (capacitance parameter related to double-layer capacitance) and its increase after LSV are most prominent for composition (3), related to the unique oxide composition even in the as-cast state, yielding a much higher capacitance. $T_1$ doubles for compositions (1) and (2), again referring to the pronounced changes on the surface layer. $R_1$ slightly increase for all the samples after LSV. For compositions (1) and (2), $T_2$ (capacitance parameter related to the oxide layer and the interfacial interactions between oxide and metallic glass) tends to decrease after LSV, whereas this value increases for composition (3). On the other hand, a pronounced rise in $R_2$, particularly for compositions (2) and (1), was observed. The smallest change in $R_2$ is observed for composition (3) since we did not observe a pitting behavior for this sample within the LSV range. Similar to these results, a high value of the CPE exponent $n$ would be expected for a smooth surface with small surface asperities.[60] **Table 3** summarizes the ECM results.

**Table 3**. ECM of the EIS results using $R_s(Q_1R_1(Q_2R_2))$. AS: As-sputtered, $R_s$: solution resistance, $T_1$: CPE parameter for double-layer capacitance, $R_1$: charge-transfer resistance, $T_2$: CPE parameter defining oxide and interface, $R_2$: oxide and interface resistance. The confidence of the fit is represented by $\chi^2$ and it is below $10^{-3}$ and generally within $10^{-4}$.

|  | (1) AS | (1) LSV | (2) AS | (2) LSV | (3) AS | (3) LSV |
|---|---|---|---|---|---|---|
| $R_s$ ($\Omega$ cm$^2$) | 78.9 | 82.2 | 114.5 | 356 | 18.53 | 18.82 |
| $T_1$ ($\Omega$ cm$^{-2}$) | 4.894 × 10$^{-6}$ | 9.457 × 10$^{-6}$ | 3.796 × 10$^{-6}$ | 7.725 × 10$^{-6}$ | 1.136 × 10$^{-5}$ | 3.164 × 10$^{-5}$ |
| $n_1$ (−) | 0.9636 | 0.9582 | 0.9664 | 0.9653 | 0.941 | 0.8866 |
| $R_1$ ($\Omega$ cm$^2$) | 7.434 × 10$^5$ | 7.718 × 10$^5$ | 1.348 × 10$^6$ | 1.52 × 10$^6$ | 5.475 × 10$^4$ | 5.618 × 10$^4$ |
| $T_2$ ($\Omega$ cm$^{-2}$) | 9.826 × 10$^{-6}$ | 5.217 × 10$^{-6}$ | 1.402 × 10$^{-5}$ | 1.677 × 10$^{-6}$ | 3.119 × 10$^{-5}$ | 8.219 × 10$^{-5}$ |
| $n_2$ (−) | 0.9519 | 0.9509 | 0.9671 | 0.9999 | 0.9099 | 0.8803 |
| $R_2$ ($\Omega$ cm$^2$) | 16.01 | 120.5 | 14.16 | 1257 | 11.14 | 19.03 |
| $\chi^2$ | 2.691 × 10$^{-3}$ | 5.816 × 10$^{-4}$ | 2.595 × 10$^{-4}$ | 1.601 × 10$^{-4}$ | 3.068 × 10$^{-4}$ | 1.307 × 10$^{-3}$ |

### 2.5. Surface Analysis

**Table 4** shows the inside-film composition after ion-etching for ca. 150 nm. The presence of Ti, Ge and Zr in different proportions was thereby confirmed for the AS ion-etched samples (**Figure S1-S3**, for compositions (1)-(3), respectively). There are remarkable composition differences, particularly for the Ti and Zr, between the XPS and EDX measurements. This can be because XPS is a surface-sensitive technique, recording data only from a few layers of the etched surface, while EDX records averages of the surface and through-thickness composition in the µm scale. N1s peak related to the metal nitride (N1s at ca. 396 eV) was also observed. There is an indication of carbide for the ion-etched samples corresponding to metal carbides. During PVD deposition, it has been shown that even for high-purity (≥99.9%) Zr targets, there is always a significant quantity of C and N inclusions (on the order of 1000 ppm), especially when the negative substrate bias voltage is not applied.[61] Ge3d and Zr4p peaks overlap, where the peak positions are indicated in Figure S1c, S2c, and S3c. Besides, the Zr-N peak is detected for composition (3) (Table 4).

**Figure 4a-c** depict the oxide formation on the (2)-AS surface sample. For all the compositions, there is a $TiO_2$ + $TiO_x$ mixture and $ZrO_2$. The metal peak areas of Ge3d5, Ti2p3 and Zr3d5 are much weaker than the bulk due to the native oxide formation and nitrogen and carbon impurities originating from the sputtering targets. A clear decrease in the $Ti^{3+}$ and $Ti^{4+}$ electron configurations and $ZrO_x$ for compositions (1) and (2) after LSV in Table 4 accounts for the changes in the charge transfer and capacitance parameters. The only slight increase is observed for the $Ti^{4+}$ for composition (3), which can also explain the rise in the $T_2$ parameter (in Table 3).

In order to observe the influence of minor etching on AS samples, only composition (2) was etched by 2 nm ((2)-AS 2nm-etched sample, **Figure 4d-f**). An increase in the $ZrO_x$ and $TiO_x$ peaks is observed. When compared to the (1)-LSV surface sample shown in Figure 5a-c, the (1)-LSV 2nm-etched sample shows a pronounced increase in the $TiO_x$ and $ZrO_x$, whereas there is no detectable increase in the case of Ge (**Figure 5d-f**). The compositional modifications of the (3)-LSV 2nm-etched sample are provided in **Figure S4,** where the largest amount of $TiO_x$ *and* $TiO_2$ are observed. XPS analyses of the (1)-AS surface, (3)-AS surface, and (2)-LSV surface samples are given in **Figure S5-S7**.

Compared to the C=O and C–O bonds, the ~~amount of~~ oxide content increases for the (2)-AS 2nm-etched sample (**Figure S8**). Again, the amount of the oxide peak increases dramatically along with C content for the (1)-LSV 2nm-etched and (3)-LSV 2nm-etched samples (**Figure**

**S9-S10**). In all cases, there is a remarkable drop in the C–O peak after minor etching. The O1s and C1s surface scans for the other compositions are given in **Figure S11-12**((1)-AS surface and (3)-AS surface) and **Figure S13** ((2)-LSV surface).

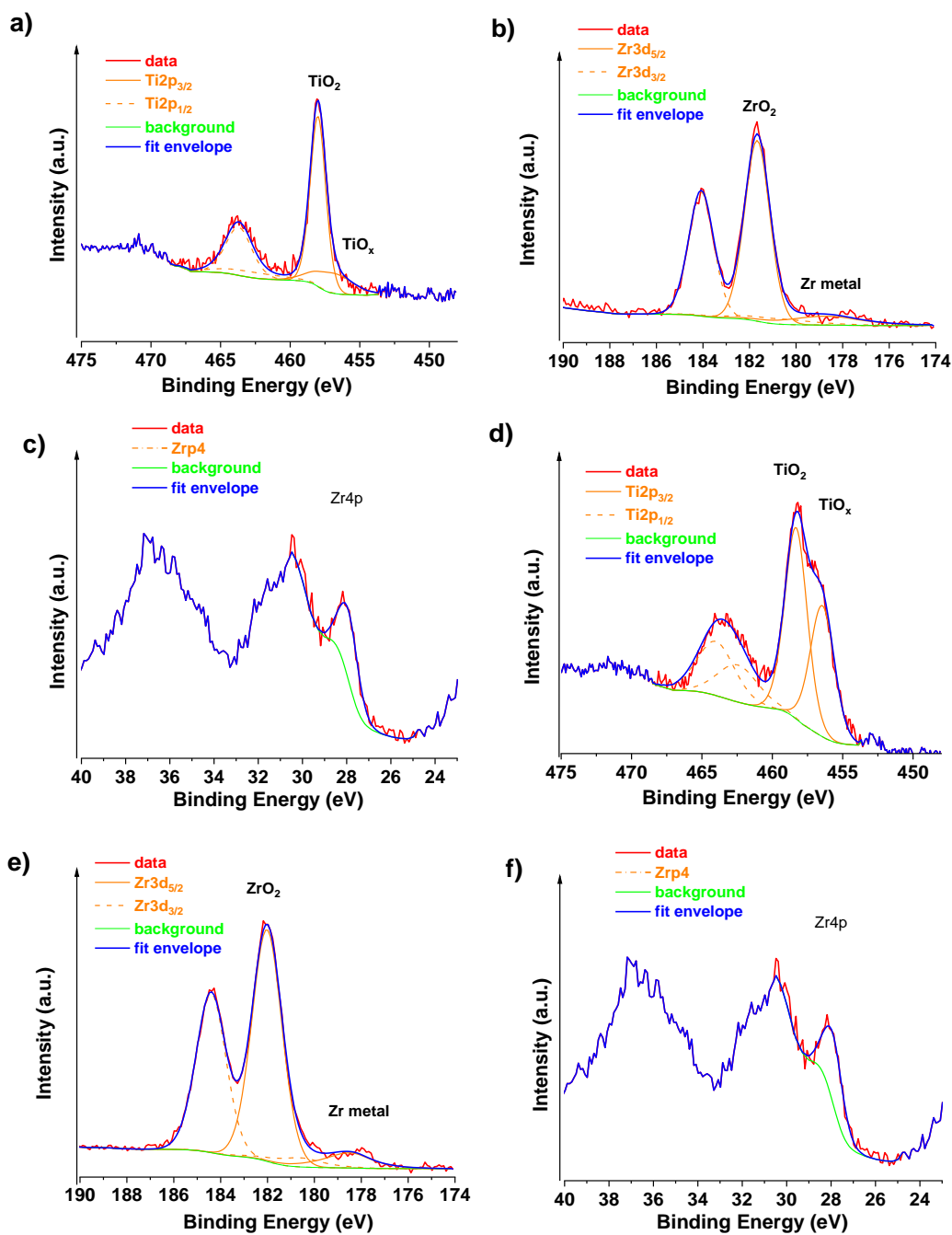

**Figure 4.** Ti2p (a), Zr3d (b), and Ge3d (c) scans of the (2)-AS surface sample and Ti2p (d), Zr3d (e), and Ge3d (f) scans of the (2)-AS 2 nm etched sample.

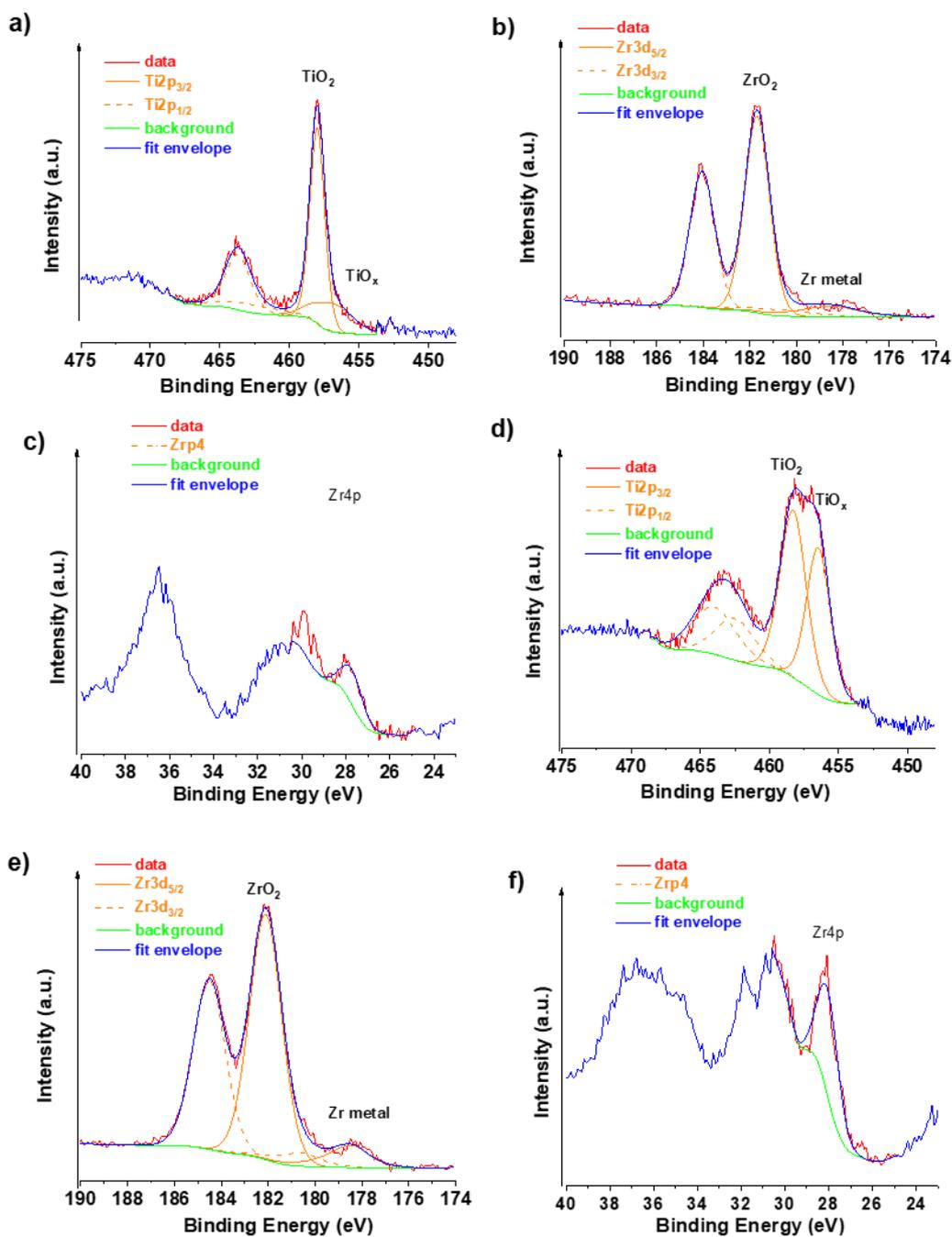

**Figure 5.** Ti2p (a), Zr3d (b), and Ge3d (c) scans of the (1)-LSV surface sample and Ti2p (d), Zr3d (e), and Ge3d (f) scans of the (1)-LSV 2 nm etched sample.

At cathodic potentials and in PBS electrolyte pH conditions, due to the possibility of ZrOH and TiOH formation, depletion of the oxide layer is promoted, which explains the decrease in the oxide content obtained from XPS. The hydroxide formation can lead to pitting corrosion, as observed in the anodic part of the LSV, which causes apparent changes in the resistance and capacitance of the compositions (1) and (2). Meanwhile, the residual carbon on the surface from the PVD process can form insoluble carbonyl groups with the native oxide or the newly formed oxide layer (at ca. 0.7 V from the reverse LSV scan) after exceeding the passivation potential, which can also contribute to the changes in the capacitance.

**Table 4**. XPS analysis of the as-spun (AS) ion-etched and AS and LSV surface compositions.

| Sample Name | Surface chemical composition (at.%) | | | | | | |
| --- | --- | --- | --- | --- | --- | --- | --- |
| | C1s | O1s oxide/C=O/ C–O | Ge3d | Ti2p metal/Ti$^{3+}$/ Ti$^{4+}$ | Zr3d met/oxide/ nitride | N1s C–N/ nitrides | Si2p/ Na1s |
| (1)-AS 150nm-etched | 0.9* | —/—/— | 28.5 | 46.5/—/— | 22.8/—/— | —/1.3 | —/— |
| (1)-AS surface | 3.7 | 48.1/12.0/1.4 | — | —/9.1/11.5 | 1.3/9.8/— | 1.7/0.9 | 0.5/— |
| (1)-LSV surface | 41.1 | 23.0/11.8/3.3 | — | —/3.0/7.9 | 0.6/4.2/— | 1.4/2.2 | 1.5/— |
| (1)-LSV 2nm-etched | 3.7 | 48.1/12.0/1.4 | — | —/9.1/11.5 | 1.3/9.8/— | 0.6/2.0 | 0.5/— |
| (2)-AS 150nm-etched | 0.9* | —/—/— | 28.3 | 46.9/—/— | 23.0/—/— | —/0.8 | —/— |
| (2)-AS surface | 51.0 | 20.1/11.2/1.9 | — | —/1.8/6.1 | 0.5/3.6/— | 1.2/1.9 | 0.7/— |
| (2)-AS 2nm-etched | 14.4 | 45.4/7.6/0.1 | — | —/8.0/10.9 | 1.0/8.6/— | 0.8/2.9 | 0.3/— |
| (2)-LSV surface | 74.0 | 5.6/11.0/0.9 | — | —/0.4/2.1 | 0.3/1.2/— | 1.5/0.8 | 1.5/0.9 |
| (3)-AS 150nm-etched | 1.1* | —/—/— | 25.5 | 61.3/—/— | 8.7/—/1.8 | —/1.6 | —/— |
| (3)-AS surface | 43.8 | 23.8/11.8/1.5 | — | —/2.7/10.2 | 0.4/1.7/— | 1.3/1.4 | 1.1/— |
| (3)-LSV surface | 37.7 | 26.7/11.9/2.1 | — | —/2.1/12.8 | 0.5/1.9/— | 1.9/1.7 | 0.7/— |
| (3)-LSV 2nm-etched | 2.8 | 40.9/17.1/1.8 | — | —/11.9/16.4 | 0.7/4.0/— | 0.7/3.3 | 0.4/— |

*C1s at ca 281.2 eV corresponding to metal carbides

## 3. Conclusions

In this work, we show for the first time that Ti-Zr-Ge metallic glass nanofilms free-from toxic and corrosive elements, i.e., Cu, Al and Ni, can be fabricated by DC magnetron sputtering. The findings show the importance of compositional adjustment to tailor the properties of the thin films. The broad diffuse X-ray diffraction maximum observed for all the samples without additional peaks confirms the fully amorphous state. The thermal properties such as relaxation, glass transition and crystallization temperatures were registered with high accuracy and sensitivity via flash differential scanning calorimetry. Ti$_{68}$Zr$_8$Ge$_{24}$ (2) shows a considerably large relaxation (284°C) and supercooled liquid (116°C) regions. A Ti-based metallic glass with such a large SCLR has been obtained for the first time and probably indicates its excellent GFA. The same composition exhibits the lowest corrosion and passivation resistance and corrosion potential, indicating its tendency for oxide formation. On the other hand, $Ti_{62}Zr_{13}Ge_{25}$ (1) has a relatively larger passive region ($\eta_{pit}$ = 1.323 V) followed by (2) ($\eta_{pit}$ = 1.213 V), indicating their high stability at low currents compared to the already discovered Ti-based metallic glasses tested in PBS solution at 37 °C. Frequency-dependent EIS measurements revealed distinct differences between compositions before and after LSV in PBS solution. The Bode magnitude of $Ti_{68}Zr_8Ge_{24}$ (2) has the largest |Z| of 195 Ω cm$^2$, and stability is reached earlier than for the other compositions at ~400 Hz. Compositions (1) and (2) with a maximum Phase angle of 86.2° and 85.9°, respectively, exhibit supercapacitive behavior in PBS solution. $R_2$ of the R(RQ(RQ)) circuit model defining the oxide layer and the interfacial interactions between oxide and metallic glass increases by almost two orders of magnitude for (2) after LSV. The influence of LSV on $Ti_{75}Zr_4Ge_{21}$ (3) is much less evidenced by minor changes in the ECM parameters. The XPS analysis confirmed that ~2 nm surface etching reveals the oxide layer composed of $TiO_2$, $TiO_x$ and $ZrO_x$. Furthermore, depletion of the oxide layer on the surface is due to the formation of the –OH groups at the cathodic potentials of the linear sweep voltammetry. For compositions (1) and (2), the increase in the O1s C=O and O1s C–O signals after LSV could be one of the main reasons behind the pronounced decrease of the interface CPE parameter $T_2$.

## 4. Methods

### 4.1. Synthesis of Thin Films by DC Magnetron Sputtering

The Ti-Ge-Zr metallic glass nanofilms (hereafter, TiNFs) were DC (direct current)-magnetron sputtered onto 350 µm thick, 20 × 7 mm$^2$ (100) Si substrates (B-doped, $\rho$ = 1–20 Ω cm) using a custom-built deposition system equipped with three magnetrons focused on a rotatable substrate holder. Zr is an ideal second element due to the zero heat of mixing with Ti along with its biocompatible nature. Ge element was chosen due to the large size and heat of mixing differences with Zr and Ti, as well as its lower Young's modulus than Si. Before coating, the substrates were sonicated in ethanol for 5 minutes and fixed to the substrate holder by Kapton® tapes. The sputtering targets (50.8 mm in diameter) were purchased from HMW Hauner, Germany, with purities of Ti: 99.995%, Zr: 99.95%, and Ge: 99.999%. One minute of pre-sputtering was applied for each target at sputter powers of Ti: 32 W, Zr: 30 W, Ge: 40 W. For film deposition, the sputter power was adjusted for each composition as follows: $Ti_{62}Zr_{13}Ge_{25}$ (Sample 1)– Ti: 140 W, Zr: 24 W, Ge: 20 W , $Ti_{68}Zr_8Ge_{24}$ (Sample 2)–

Ti: 140 W, , Zr: 40 W, Ge: 23 W, Ti$_{75}$Zr$_4$Ge$_{21}$ (Sample 3)– Ti: 140 W, Zr: 12 W, Ge: 17 W. The base pressure was between 8.3 × 10$^{-4}$ – 7.5 × 10$^{-5}$ Pa and the Ar flow was set to 30 sccm resulting in a constant process pressure of 0.5 Pa. Sputtering was performed at room temperature, without substrate etching or biasing. The sputtering time was between 13 to 17 min, where the final film thicknesses measured by laser interferometry are 465 nm for the Ti$_{62}$Zr$_{13}$Ge$_{25}$ and Ti$_{68}$Zr$_8$Ge$_{24}$ and 420 nm for the Ti$_{75}$Zr$_4$Ge$_{21}$ composition, corresponding to growth rates of ~35 nm min$^{-1}$.

### 4.2. X-ray Diffraction

Structural characterization of TiNFs was performed using a Rigaku SmartLab 5-axis X-ray diffractometer with Cu K$\alpha$ radiation and Bragg-Brentano $\theta$-$2\theta$ configuration. The grazing incidence mode was employed with high surface sensitivity (incident angle 2°). For the first broad diffraction peak, the Pseudo Voigt function, a convolution of Gaussian and Lorentzian peak fit, was utilized.

### 4.3. Flash Differential Scanning Calorimetry

Thermophysical characterization was conducted using a Mettler-Toledo Flash DSC 2. The initial sample temperature of the Flash-DSC was brought to 233 K with the aid of a Huber intracooler TC90. The FDSC probes were prepared by slicing tiny TiNFs of ~1 mm average diameter using a scalpel under a stereomicroscope and subsequently transferring these pieces using a microsurgical tweezer onto the active center of a conditioned and temperature-corrected MultiSTAR UHS1 sensor. An Ar flow of 80 ml min$^{-1}$ was applied during the measurement to protect the sample from oxidation. The temperature was first raised at a rate of 250 °C min$^{-1}$ of up to 980 °C and subsequent cooling to room temperature. Thermophysical analysis was performed using the STARe excellence thermal analysis software.

### 4.4. Electrochemical Polarization and Impedance Spectroscopy

A premixed phosphate buffer solution (PBS) (ROTI®Cell PBS - NaCl: 154.004 mM, Na$_2$HPO$_4$: 5.599 mM, KH$_2$PO$_4$: 1.058 mM) with a pH of 7.4 ± 0.1 was used for this study. For each measurement, 5 mL solution was deaerated for 15 min using Ar gas. One edge of the TiNF was coiled by copper tape (Busch 1799) to fit into alligator clips (SKS Hirschmann KLEPS 2600), holding the working electrode and establishing high electrical conductivity. The samples were immersed in the solution preheated to 37 ± 1 °C and kept for 1 h until reaching stability. The electrochemical measurements were conducted using a Pt counter electrode (0.8 mm diameter) and Ag(s) /AgCl(s) with saturated KCl solution with a redox potential of +0.640 V ($E_{Ag/AgCl}$ + $E°_{Ag/AgCl}$ (0.197 – 0.00101× $T$ (37–25)) V + 0.0615 × pH (7.4)) vs. a reference hydrogen electrode.[62] The submerged surface areas of the samples were 0.425 cm$^2$ ($Ti_{62}Zr_{13}Ge_{25}$), 0.740 cm$^2$ ($Ti_{68}Zr_8Ge_{24}$) and 0.281 cm$^2$ ($Ti_{75}Zr_4Ge_{21}$). A micro-cell kit with a conical solution reservoir was used to establish closer positioning of the three-electrode system for very high accuracy. The difference in submerged areas is mainly due to the width of the substrates and difficulty in the visual arrangement of such small samples. The samples were polarized between –1 V and 0.8 V using cyclic voltammetry for two cycles to establish a stable solution-electrode interface. Subsequently, polarization using linear sweep voltammetry (LSV) was performed at a scan rate of 0.05 V s$^{-1}$ from –1 V to 0.8 V followed by a reverse cycle until the repassivation point using a current range of 200 μA and $E$ (potential) and $I$ (current) filters of 100 Hz. The electrochemical measurements were performed using a PARSTAT 4000A Potentiostat Galvanostat (Princeton Applied Research, USA) equipped with a VersaStudio 2.62.2 software module. The Tafel fit function of this software was utilized to measure the cathodic $\beta_c$ and anodic $\beta_a$ Tafel fits, corrosion current density, $j_{corr}$, and corrosion potential, $E_{corr}$. Electrochemical impedance spectroscopy (EIS) studies were conducted at open circuit potential (OCP) at an AC amplitude of 10 mV recorded from 100000 Hz to 0.1 Hz before and after the polarization study. The simulation of the EIS data was conducted with an electrical equivalent circuit of R(QR(QR)) using the ZSimpWin V.3.10 analysis program.

### 4.5. Morphology and Composition Analysis

Morphological analysis was performed using a field-emission scanning electron microscope (FESEM, Zeiss Leo 1525) at an acceleration voltage of 5 kV for imaging and 30 kV to determine chemical composition. The composition study was conducted by a Bruker XFlash Detector 6|60 energy-dispersive X-ray spectroscopy (EDS) unit. The theoretical densities of the selected compositions, i.e. $\rho$ = 5.235 g cm$^{-3}$ ($Ti_{62}Zr_{13}Ge_{25}$), $\rho$ = 5.066 g cm$^{-3}$ ($Ti_{68}Zr_8Ge_{24}$), $\rho$ = 4.903 g cm$^{-3}$ ($Ti_{75}Zr_4Ge_{21}$), and film thickness values were included for the accurate estimation of the compositions. $\rho$ was calculated from the equation $\frac{1}{\rho_{MG}} = \frac{\omega_{Ti}}{\rho_{Ti}} + \frac{\omega_{Zr}}{\rho_{Zr}} + \frac{\omega_{Ge}}{\rho_{Ge}}$. Here, $D_{Ti}$, $D_{Zr}$ and $D_{Ge}$ are the densities and $\omega_{Ti}$, $\omega_{Zr}$ and $\omega_{Ge}$ are the mass fractions in the mixture. The mean composition was calculated from measuring 12 different locations as: sample (1) – Ti: 62.60 ± 0.16, Zr: 12.65 ± 0.12, Ge: 24.75 ± 0.16, sample (2) – Ti: 68.24 ± 0.13, Zr: 7.73 ± 0.07, Ge: 24.03 ± 0.08, sample (3) – Ti: 74.80 ± 0.17, Zr: 4.00 ± 0.15, Ge: 21.20 ± 0.14.

### 4.6. Tribological Analysis

Atomic force microscopy (AFM) imaging of the TiNF surfaces was performed with a Dimension 3100 scanning probe microscope (Veeco, USA) in tapping mode using a standard silicon tapping mode cantilever with a nominal tip radius of 10 nm. The scan size was $5 \times 5$ $\mu m^2$, and the scan rate was 1 Hz. The real surface area was approximated by the triangulation method. Root-mean-square roughness and average height were determined using the statistical quantities tool of the Gwyddion – Free SPM data analysis software.[63]

### 4.7. X-ray Photoelectron Spectroscopy

X-ray photoelectron spectroscopy (XPS) analyses were implemented by a Thermo Scientific K-Alpha compact XPS system (Thermo Fisher Scientific, UK) attached to a micro-focused, monochromatic Al Kα X-ray source (1486.68 eV). The spectra were acquired in the constant analyzer energy mode with the pass energy of 200 eV for the survey. Narrow regions were collected with the pass energy of 50 eV. Charge compensation was achieved with the system Ar flood gun. Argon sputter cleaning for 10 s (etching rate of ca 0.2 nm/s ~ 10 nm) and 300 s (etching rate of ca 0.5 nm/s ~ 150 nm) was performed with an Ar ion gun. Thermo Scientific Avantage software, version 5.9929 (Thermo Fisher Scientific), was used for digital acquisition and data processing. Spectral calibration was determined using the automated calibration routine and the internal Au, Ag and Cu standards supplied with the K-Alpha system. The surface compositions (in atomic %) were determined by considering the integrated peak areas of detected atoms and the respective Scoffield sensitivity factor.

## Supporting Information

Supporting Information is available from the Wiley Online Library or from the author.

## Acknowledgements

The authors thank A. Sezai Sarac for the fruitful discussions on electrochemical data interpretation and Parthiban Ramasamy during composition selection. B.S. acknowledges the Austrian Science Fund (FWF) under project grant I3937-N36. M.M. and M.O. acknowledge the support from VEGA project 02/0006/22. This work was supported by the Fundamental Research Funds for the Central Universities (No. NS2019034) and was performed during the implementation of the project Building-up Centre for advanced materials application of the Slovak Academy of Sciences, ITMS project code 313021T081 supported by the Research & Innovation Operational Programme funded by the ERDF, and the National Key Research and Development Program (No. 2019YFE0107000), innovation program for Marie Skłodowska-Curie Individual Fellowship, with the acronym "MAGIC" and grant agreement N. 892050.

## Conflict of Interest

The authors declare no conflict of interest.

## Data Availability Statement

The data that support the findings of this study are available from the corresponding author upon reasonable request.

**Supplementary Data**

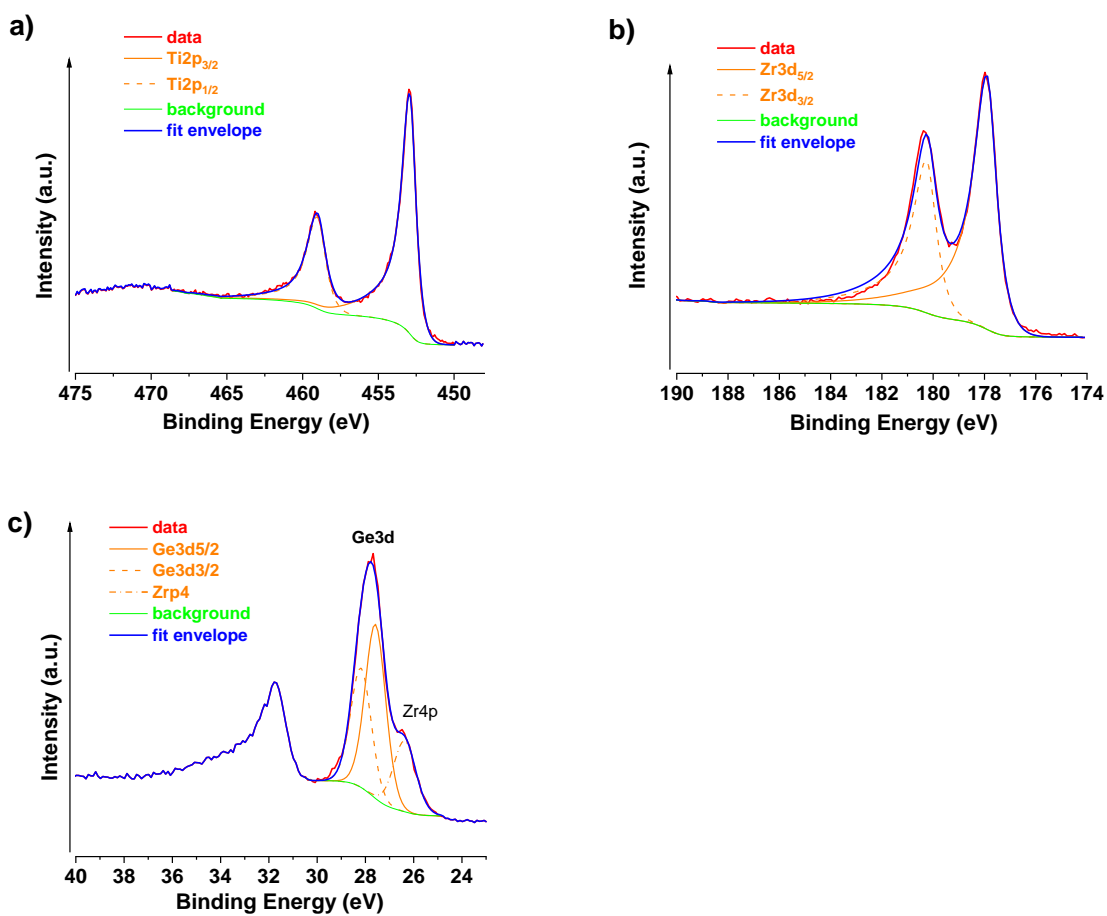

**Figure S1**. (a) Ti2p, (b) Zr3d, and (c) Ge3d scans of the (1)-AS ion-etched sample.

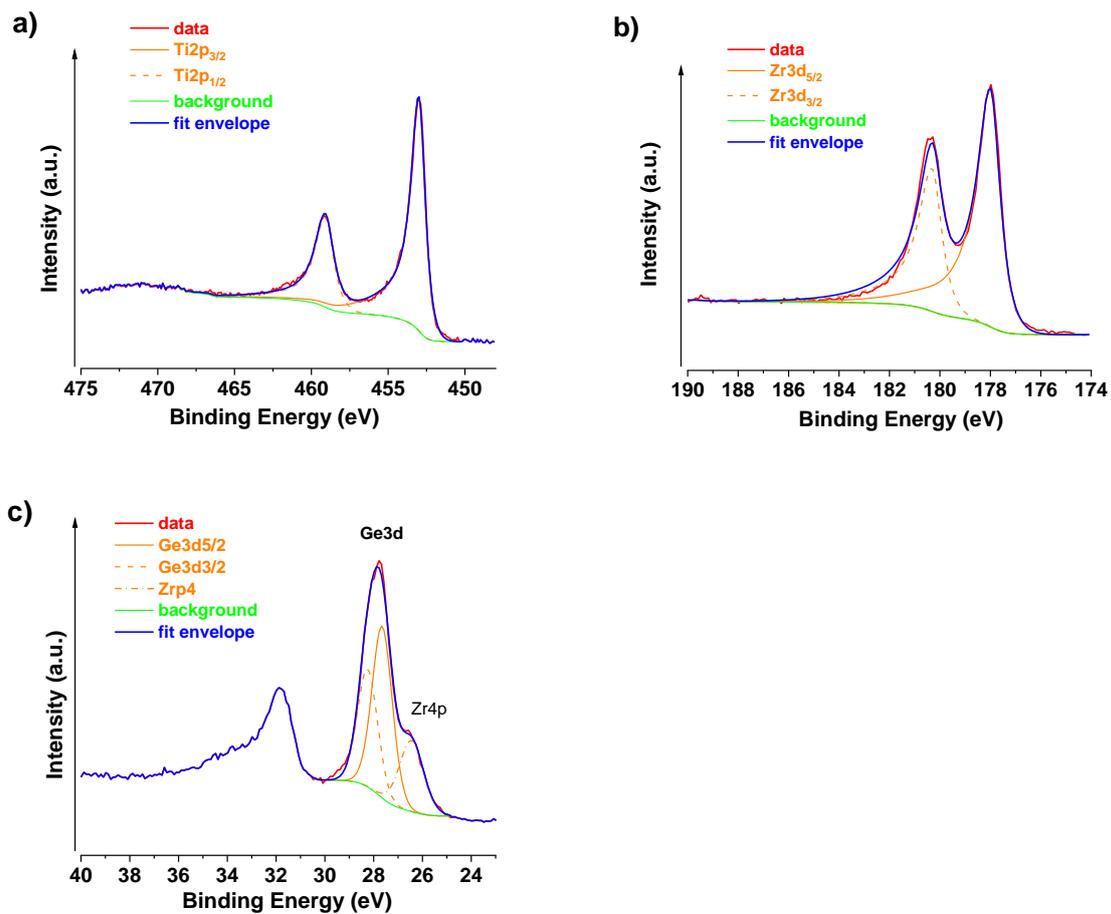

**Figure S2**. (a) Ti2p, (b) Zr3d, and (c) Ge3d scans of the (2)-AS ion-etched sample.

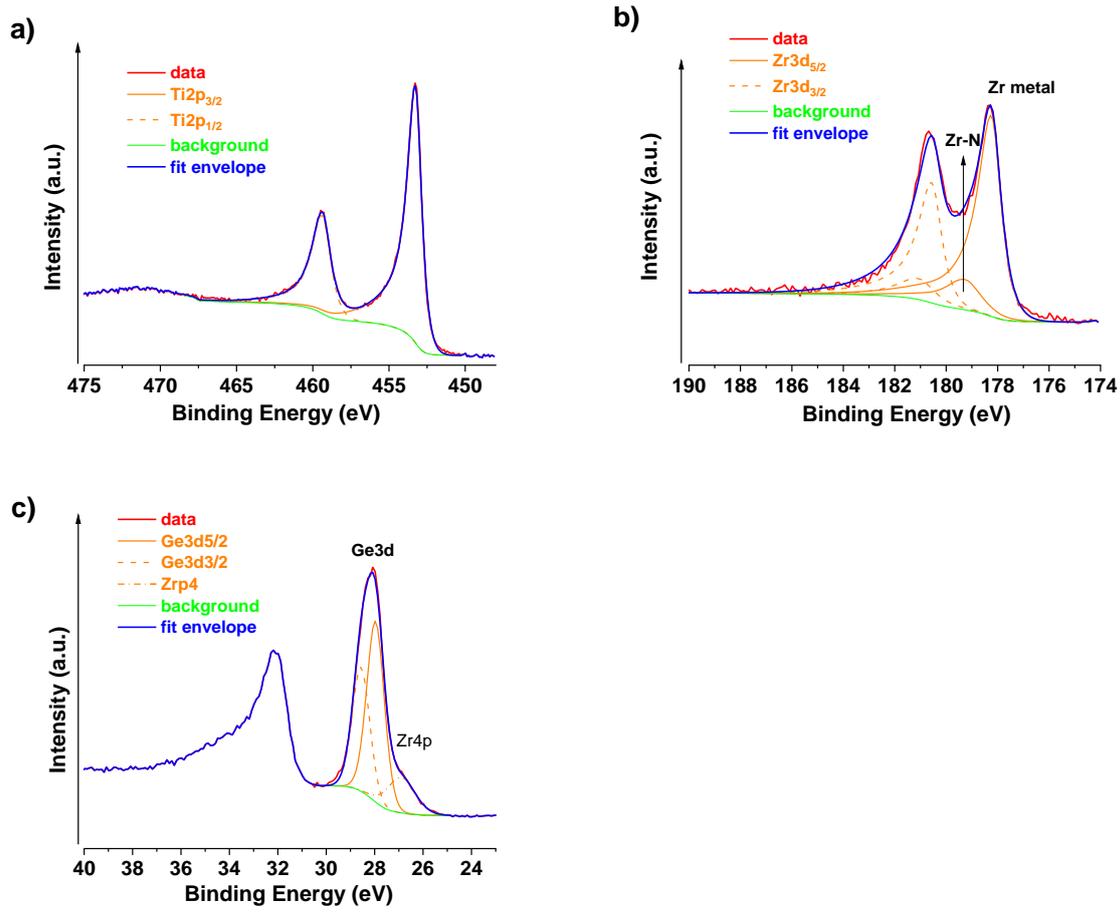

**Figure S3**. Ti2p (a), Zr3d (b), and Ge3d (c) scans of the (3)-AS ion-etched sample.

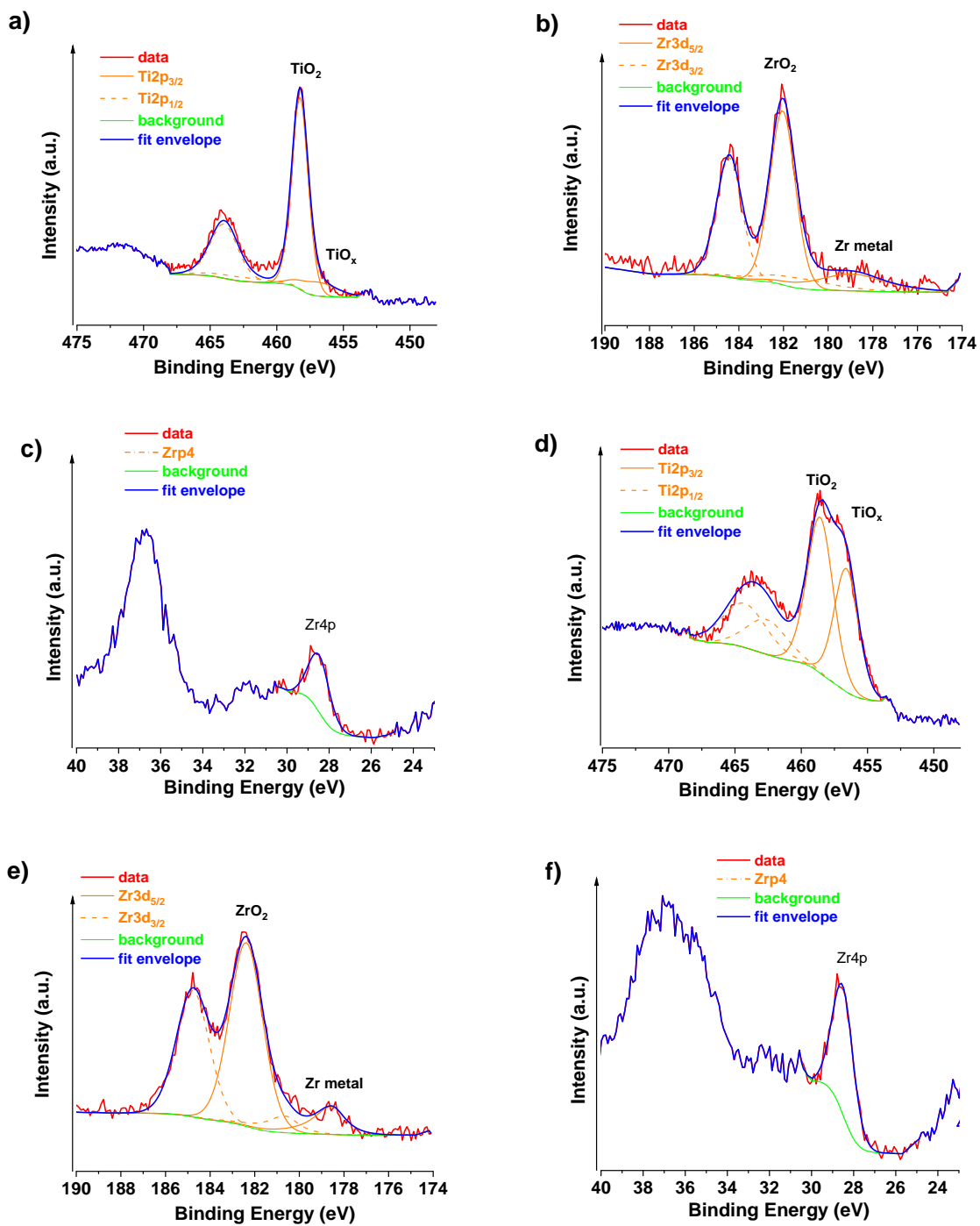

**Figure S4.** Ti2p (a), Zr3d (b), and Ge3d (c) scans of the (3)-LSV surface sample and Ti2p (d), Zr3d (e), and Ge3d (f) scans of the (3)-LSV 2 nm etched sample.

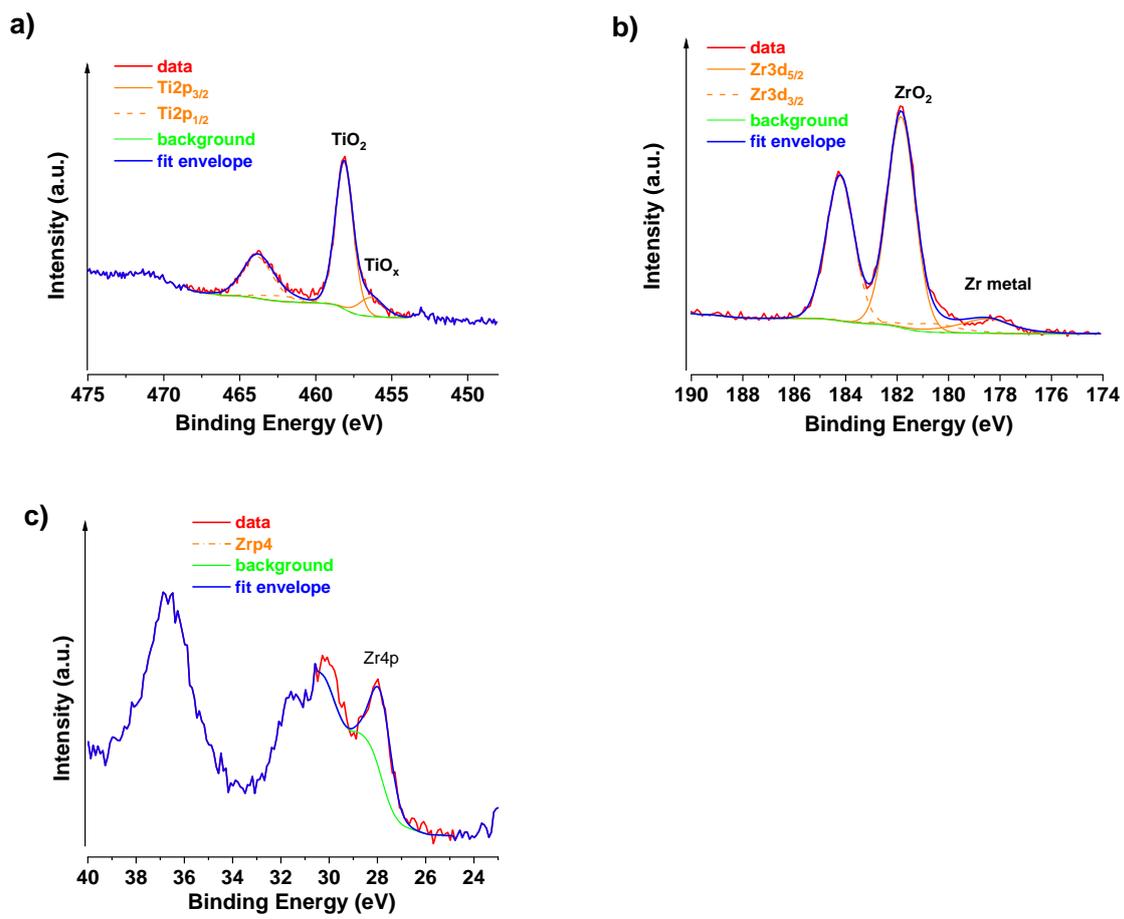

**Figure S5.** Ti2p (a), Zr3d (b), and Ge3d (c) scans of the (1)-AS surface sample.

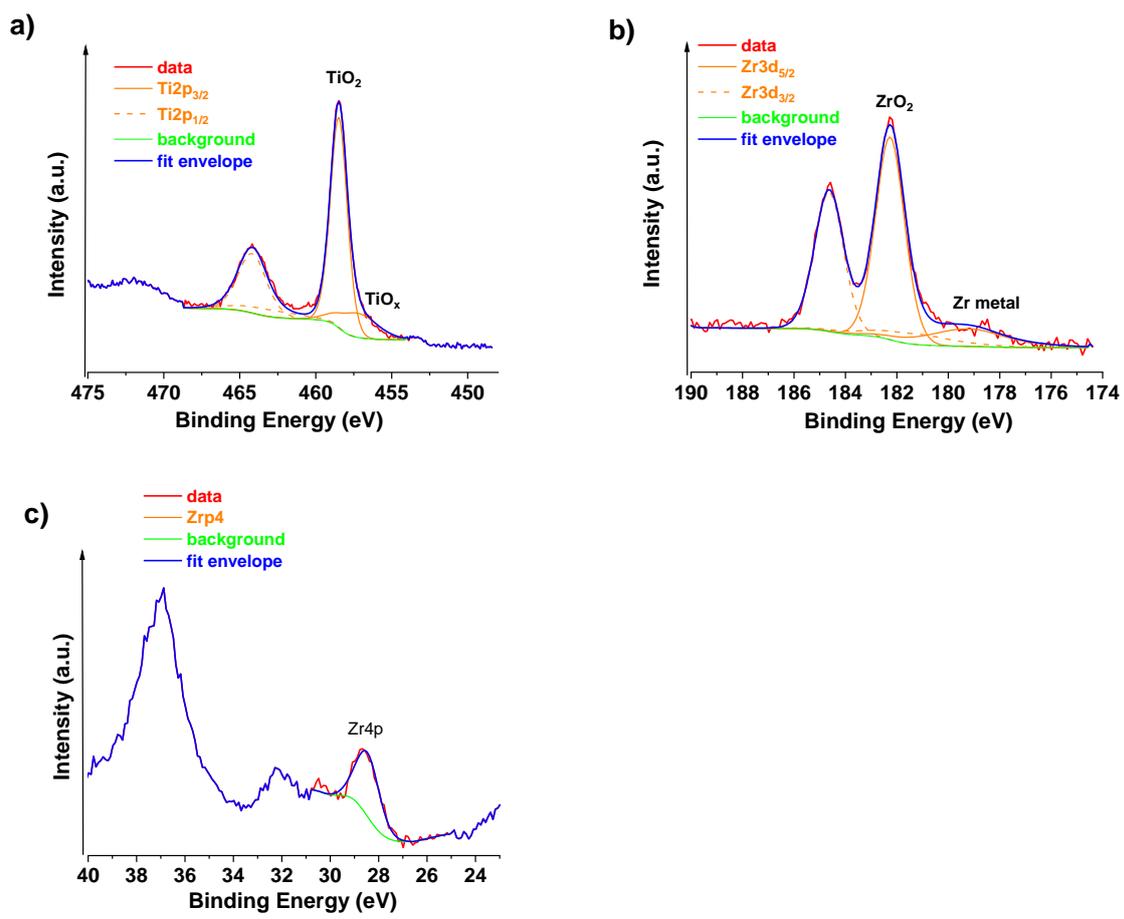

**Figure S6**. Ti2p (a), Zr3d (b), and Ge3d (c) scans of the (3)-AS surface sample.

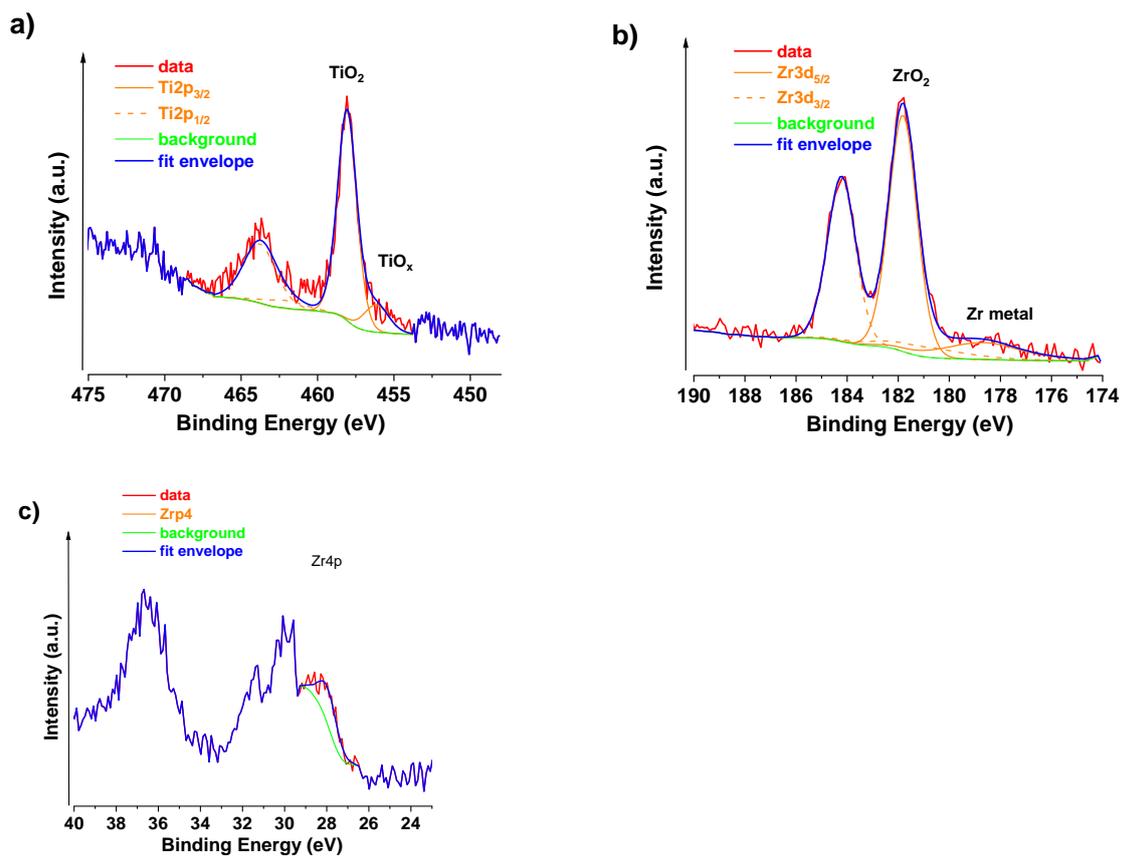

**Figure S7.** Ti2p (a), Zr3d (b), and Ge3d (c) scans of the (2)-LSV surface sample.

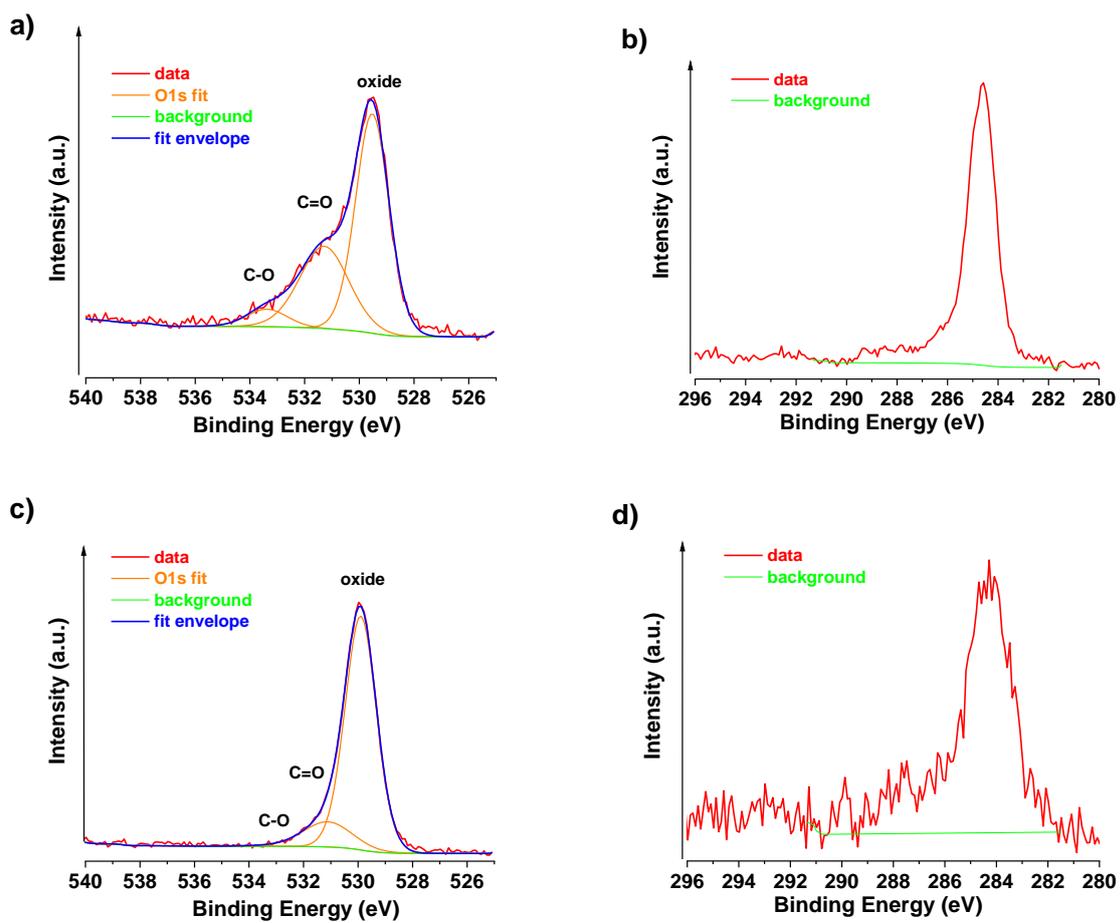

**Figure S8**. O1s (a) and C1s (b) scans of the (2)-AS surface sample and O1s (c) and C1s (d) scans of the (2)-AS 2 nm etched sample.

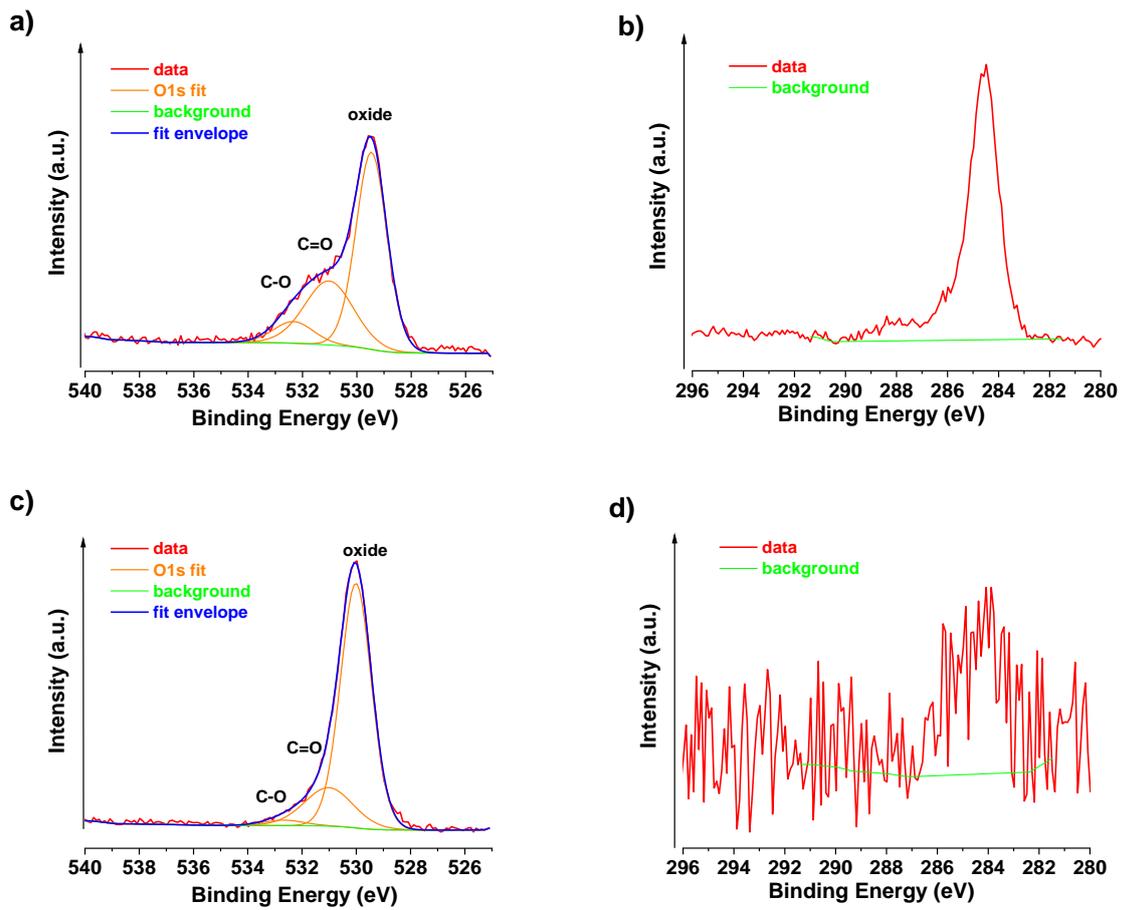

**Figure S9.** O1s (a) and C1s (b) scans of the (1)-LSV surface sample and O1s (c) and C1s (d) scans of the (1)-LSV 2 nm etched sample.

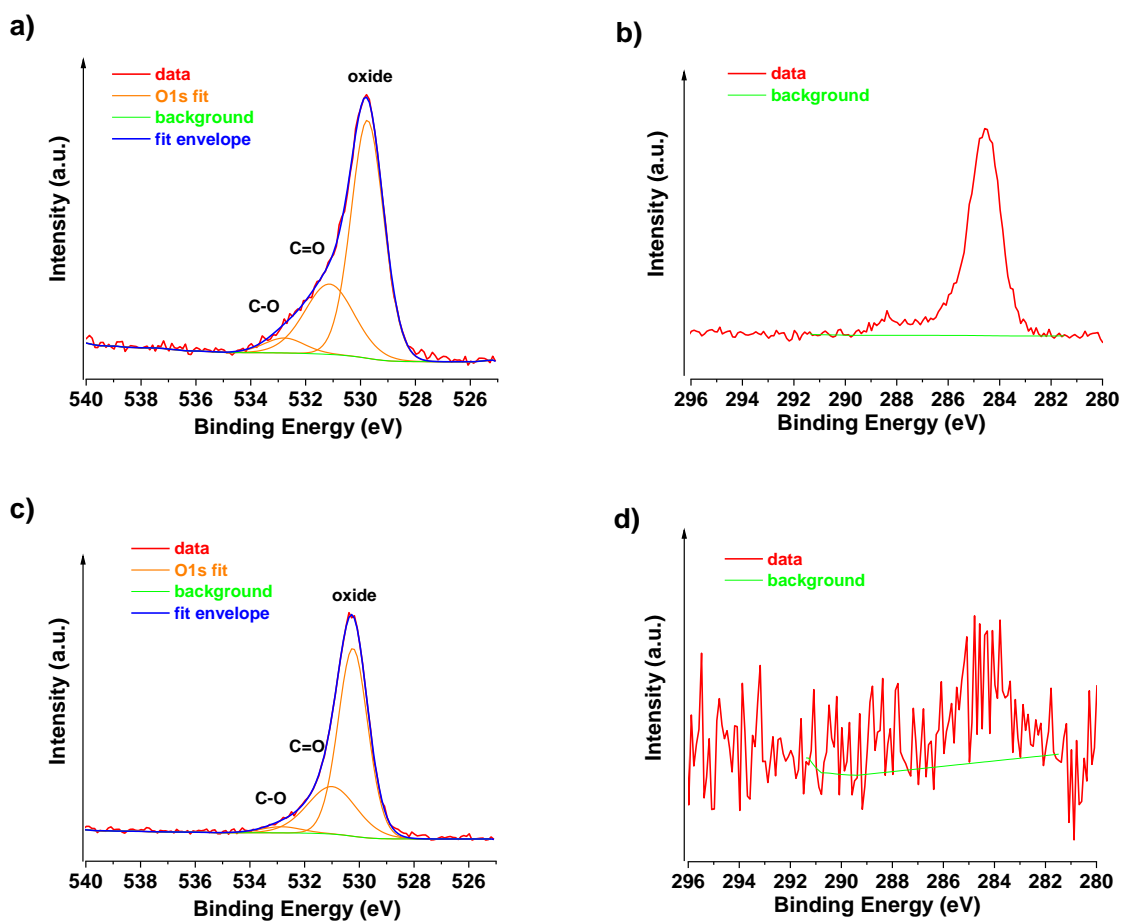

**Figure S10**. O1s (a) and C1s (b) scans of the (3)-LSV surface sample and O1s (c) and C1s (d) scans of the (3)-LSV 2 nm etched sample.

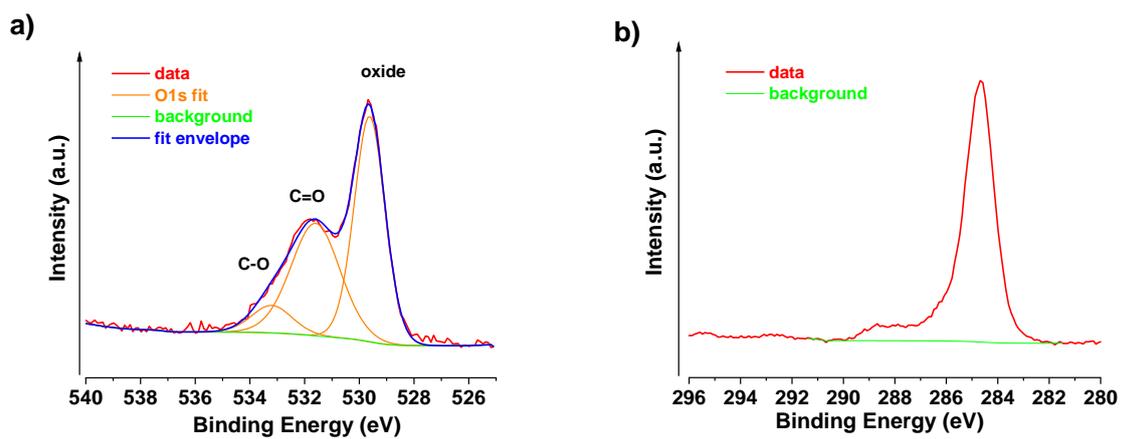

**Figure S11.** O1s (a) and C1s (b) scans of the (1)-AS surface sample.

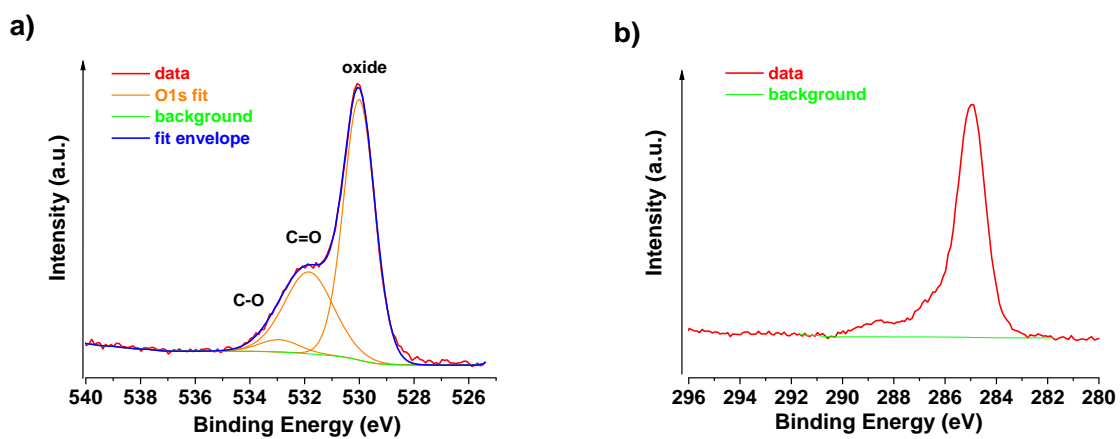

**Figure S12.** O1s (a) and C1s (b) scans of the (3)-AS surface sample.

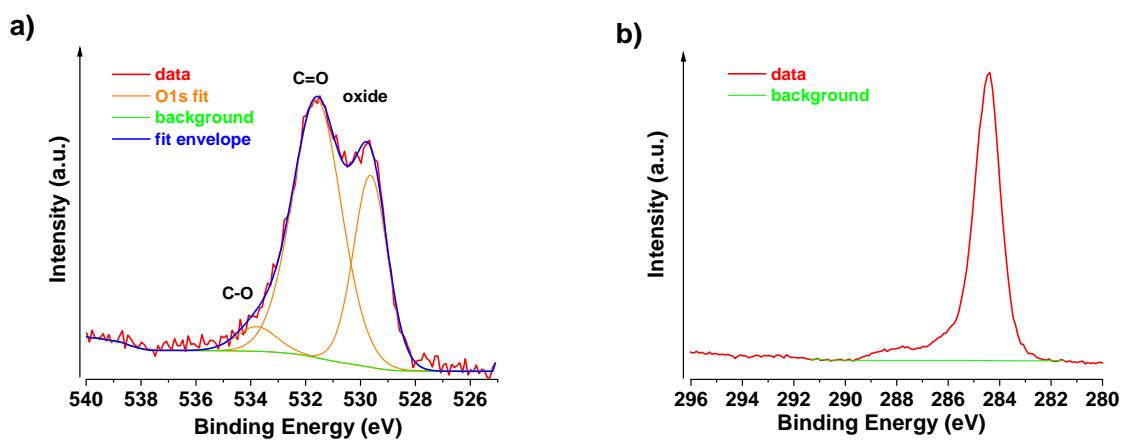

**Figure S13.** O1s (a) and C1s (b) scans of the (2)-LSV surface sample.

**For Table of Contents Only**

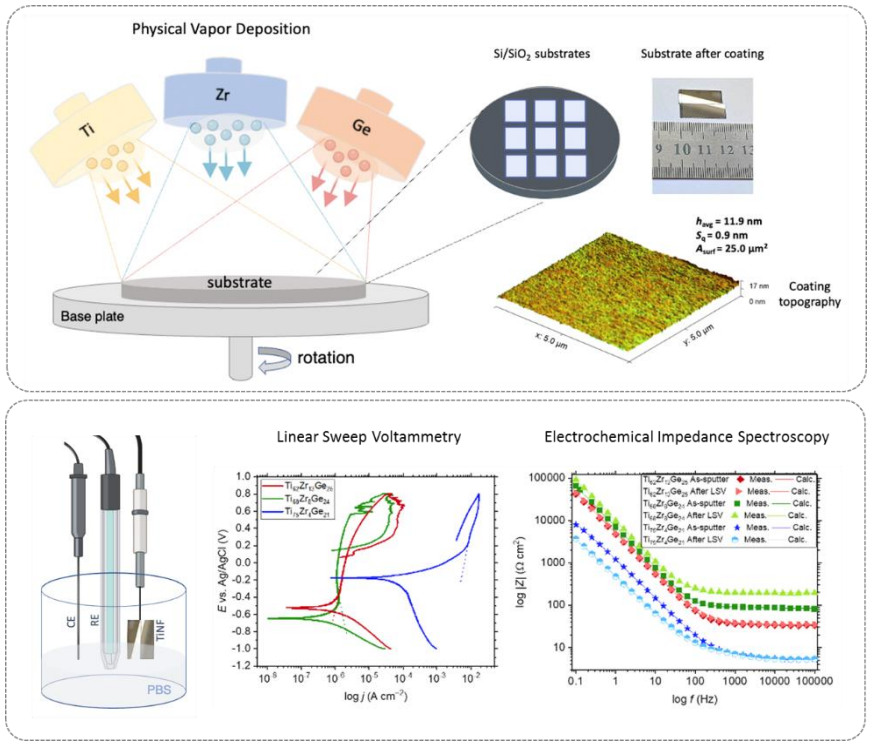

Composition of DC magnetron sputtered Ti-Zr-Ge metallic glass nanofilms can be tailored to achieve enhanced biocorrosion resistance, high passivation and supercapacitive behavior within PBS solution. Depletion of the oxide layer on the surface is due to the formation of –OH groups at the cathodic potentials, where an increase in O1s C=O and O1s C–O signals after LSV is observed.